\documentclass[prm,aps,a4paper,twocolumn,floatfix,repreprint,superscriptaddress,longbibliography]{revtex4}
\usepackage{amsmath}
\usepackage{amsfonts}
\usepackage{graphicx}
\usepackage{upgreek}
\usepackage{bm}
\usepackage{color}
\usepackage{multirow}
\usepackage{tabularx}
\usepackage{natbib}
\usepackage{hyperref}
\usepackage{caption}
\usepackage{subcaption}
\usepackage{graphicx}
\usepackage[normalem]{ulem}
\usepackage{relsize}

\def\be{\begin{equation}}
\def\ee{\end{equation}}
\DeclareUnicodeCharacter{2212}{-}
\begin{document}

\title{Magnetic structure and electronic properties of mixed-metal Ruddlesden-Popper oxide LaSrCo$_{1/2}$Fe$_{1/2}$O$_4$}

\author{Dina I. Fazlizhanova}
\affiliation{Skolkovo Institute of Science and Technology, Moscow 121205, Russia}
\author{Ruslan G. Batulin }
\affiliation{Kazan (Volga Region) Federal University, Kremlevskaya st., 18, 420008 Kazan, Russia}
\author{Tatiana I. Chupakhina}
\affiliation{Institute of Solid State Chemistry of the Russian Academy of Sciences (UB), Pervomaiskaya St., 91, Ekaterinburg, 620990, Russia}
\author{Yulia.A. Deeva }
\affiliation{Institute of Solid State Chemistry of the Russian Academy of Sciences (UB), Pervomaiskaya St., 91, Ekaterinburg, 620990, Russia}
\author{Sergey V. Levchenko}
\affiliation{Skolkovo Institute of Science and Technology, Moscow 121205, Russia}


\begin{abstract}
Atomic, electronic, and magnetic structure of LaSrCo$_{1/2}$Fe$_{1/2}$O$_4$ mixed-metal Ruddlesden-Popper oxide is investigated theoretically using self-consistent ACBN0 DFT + $U$ approach. We show that the electronic and magnetic properties strongly depend on the distribution of transition-metal and La/Sr ions in the lattice.  Fe-Fe exchange is found to be antiferromagnetic, whereas Co-Co and Fe-Co exchange is ferromagnetic. We find that Co spin states depend on the distribution in both La/Sr and transition-metal ions. The most energetically favorable configuration is ferromagnetic, whereas the majority of metastable configurations are antiferromagnetic. LaSrCo$_{1/2}$Fe$_{1/2}$O$_4$ was synthesized using spray-pyrolysis method. Magnetization measurements revealed  antiferromgnetic behavior, thus indicating presence of metastable configurations in the synthesized material. In addition, we show that in complex systems, such as mixed-metal Ruddlesden-Popper phases, Hubbard $U$ values have to be determined for each crystallographically non-equivalent site.

\end{abstract}

\maketitle	

\section{Introduction}
In recent decades, there has been significant interest in the hydrogen economy as a result of increasing energy demand and environmental concerns related to the use of fossil fuels. One potential method for producing environmentally friendly hydrogen is through water electrolysis. However, the slow kinetics of the oxygen evolution reaction (OER) restricts the efficiency of this process. OER is also an important process for fuel cells and metal-air batteries. 

Ruddlesden-Popper (RP) oxides have shown great potential as catalysts for OER \cite{LaSrNiFeO4}. These oxides have a general formula A$_{1+n}$B$_n$O$_{3n+1}$, where A represents an alkali or alkali earth element, B represents a transition metal, and $n$ denotes the number of perovskite layers ABO$_3$ separated by a rock-salt layer AO. The ability to adjust their properties by substituting different elements in the A and B sites makes RP oxides attractive for the rational design of cost-effective, environmentally friendly, and stable catalysts for OER.

Several attempts have been made in literature to identify descriptors that can be used to predict the activity of catalysts. For instance,  O 2$p$-band center \cite{O2p, descriptor}, transition-metal's band overlap \cite{LaSrNiFeO4}, hybridization of transition metal 3$d$ and oxygen  2$p$ electrons \cite{3d_2p_overlap}, $d$-band center relative to O 2$p$-band center \cite{charge_transfer}, unoccupied 3$d$-band center, Metal−O−Metal bond angle, tolerance factor \cite{statistical_analysis}.  However, these properties could depend on the distributions of dopants. 
Very often descriptors are derived from density-functional theory (DFT) calculations. This approach has several drawbacks. First, in order to perform DFT calculations, one needs to create a supercell to accommodate substitutions, or use virtual atoms. Second, standard DFT calculations correspond to temperature 0 K, so that effects such as magnetic and/or charge disorder are not taken into account.

It is generally believed that the most favorable configuration of atoms in A and B sublattices of A$_{2-x}$A'$_x$B$_{1-y}$B'$_y$O$_{4}$, which defines the electronic and other properties of the crystal, is the one in which the number of valence electrons is balanced first between the layers (i.e., the ratio of A and A'  is the same in each A-site layer, and the ratio of B and B' is the same in each B-site layer), and then within the layer (i.e., the number of A-O-A' and B-O-B' bridges is maximized in the compound \cite{LaSrNiFeO4}). With this assumption, the descriptors can be evaluated for each particular elemental composition. However, magnetic structure, charge ordering \cite{charge_order}, and various spin states \cite{high_spin_low_spin} make predicting the energetically favorable configuration challenging. Another approach is to consider the fully randomized configuration, assuming that the difference of descriptor values from composition to composition exceeds the differences within one composition from configuration to configuration \cite{descriptor}. However, this assumption neglects the possibility of using variation of long-range ordering in A and B sites as means for designing RP oxides with optimal properties along with elemental composition and the number of perovskite layers $n$.

In this work, we determine the influence of distribution of dopants in A and B sites on stability, lattice parameters, electronic structure, magnetic ordering, and spin state for LaSrCo$_{1/2}$Fe$_{1/2}$O$_4$ as an example, using DFT calculations and experiments. The particular choice of elemental composition is motivated by its relevance for practical applications \cite{synthesis3} and a rich electronic structure due to different possible spin states of Co$^{3+}$ ions in the lattice. Due to the delicate balance between Hund's exchange energy and the crystal-field-splitting energy, Co$^{3+}$ ions' spin state is sensitive to crystal-field interaction, which is in turn affected by the changes in internal pressure due to the chemical substitution (or external pressure and temperature). In our investigation, we systematically evaluated a range of properties for nine distinct structures featuring varied distributions of La/Sr and Co/Fe. Employing the ACBN0 DFT+$U$ self-consistent approach, we studied the influence of atom distributions in each sublattice on the properties of the RP oxide. A key aspect of our analysis involved the self-consistent determination of the Hubbard $U$ parameter for each unique atomic environment. To complement our theoretical findings, we synthesized a LaSrCo$_{1/2}$Fe$_{1/2}$O$_4$ powder sample and conducted magnetization measurements across a range of temperatures. These measurements were carried out under a magnetic field ranging from -10,000 to 10,000 Oe, providing valuable insights into the magnetic behavior of the synthesized material.

 \section{Methods} 
\begin{figure}
      \subfloat[]{\includegraphics[width=0.34\linewidth]{LaSrFe0.5Co0.5O4.pdf} }
 \subfloat[]{\includegraphics[width=0.2\linewidth]{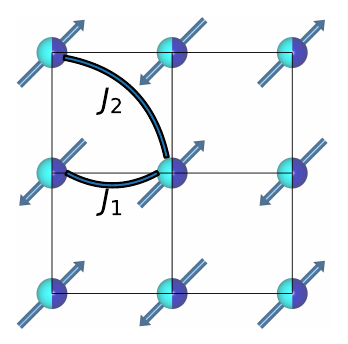}}
 \subfloat[]{\includegraphics[width=0.33\linewidth]{LaSrFe0.5Co0.5O4_AFM_A.pdf}}
	\caption{(a) The crystal structure of LaSrCo$_{1/2}$Fe$_{1/2}$O$_4$. O1 and O2 denote two nonequivalent positions of oxygen. O1 occupies the 4e crystallographic site in the perovskite layer, while O2 occupies the 4c site in the rock-salt layer. La/Sr occupies the 4e site, and Fe/Co occupies the 2a site, respectively. (b) Top view of the transition-metal layer showing the nearest- and next-nearest-neighbor spin interactions. (c) AFM-A antiferromagntic: spin directions interchanged between the layers. }
 \label{magn_struc}
 \end{figure}
        \begin{figure}
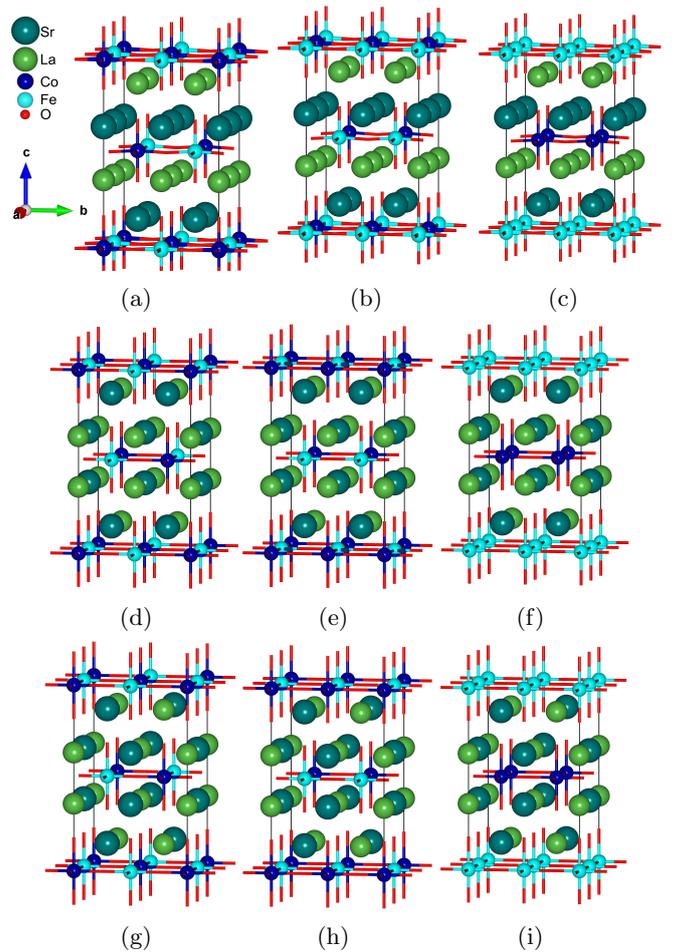

        \subfloat[]{\includegraphics[width=0.4\linewidth]{A_sym_B_w4_06_legend.pdf}}
        \subfloat[]{\includegraphics[width=0.3\linewidth]{A_sym_B_w4_04.pdf}}
        \subfloat[]{\includegraphics[width=0.3\linewidth]{A_sym_B_sym.pdf}}
        
        \subfloat[]{\includegraphics[width=0.3\linewidth]{A_w4_04_B_w4_06.pdf}}
        \subfloat[]{\includegraphics[width=0.3\linewidth]{A_w4_04_B_w4_04.pdf}}
        \subfloat[]{\includegraphics[width=0.3\linewidth]{A_w4_04_B_sym.pdf}}
        
        \subfloat[]{\includegraphics[width=0.3\linewidth]{A_w4_06_B_w4_06.pdf}}
        \subfloat[]{\includegraphics[width=0.3\linewidth]{A_w4_06_B_w4_04.pdf}}
        \subfloat[]{\includegraphics[width=0.3\linewidth]{A_w4_06_B_sym.pdf}}
	\caption{ Various distributions of La/Sr and Co/Fe considered in this work. Along the rows, the distribution of La/Sr remains the same, but there are different distributions in the Co/Fe positions. Along the columns, the distribution of Co/Fe positions remains the same, but there are different distributions of La/Sr. a-c: Layers of La and Sr are interchanged in the structure. d-f: La/Sr is distributed uniformly between the layers but ordered in stripes within the same layer. g-i: La/Sr atoms are distributed uniformly. In a, d, g: Co/Fe is distributed uniformly. In b, e, h: Co/Fe is distributed uniformly between the layers but ordered in stripes within the same layer. In c, f, i: Layers of Co and Fe are interchanged in the structure. For the sake of clarity the oxygen atoms are not shown. }
 \label{struc}
\end{figure}

\subsection{Synthesis}

We synthesized the complex oxide LaSrCo$_{1/2}$Fe$_{1/2}$O$_4$ through the pyrolysis of nitrate-organic mixtures of the corresponding components, employing the “solution combustion” method \cite{combustion_update}. “Solution combustion” is a widely utilized method for synthesizing complex oxides with a perovskite-like structure \cite{combustion_review}. The choice of fuel is crucial in this process \cite{synthesis5}. For the pyrolysis of an organic nitrate composition containing La, Sr, Co and Fe nitrates, we utilized diubstituted ammonium citrate (C$_6$H$_{14}$N$_2$O$_7$). This technique is applicable for obtaining complex oxides of various structures and compositions \cite{synthesis6, patent}.




The starting reagents were Sr(NO$_3$)$_2$ (strontium nitrate, chemically pure), La(NO$_3$)$_3\cdot6$H$_2$O (lanthanum nitrate, extra
pure), Fe(CO)$_5$ (carbonyl, specially pure), Co(NO$_3$)$_2\cdot$6H$_2$O (cobalt (II) nitrate, chemically pure), HNO$_3$ (special purity grade), C$_6$H$_{14}$N$_2$O$_7$ (reagent grade), selected as an organic component, ensuring the pyrolysis process occurs in the solution self-ignition mode.

Iron (II) nitrate is prepared by dissolving a stoichiometric amount of carbonyl iron in dilute (1:1) nitric acid under low heat. The nitrate samples in the amount of 0.01 were dissolved in distilled water. The resulting solutions were mixed. Next, a twofold excess of disubstituted ammonium citrate C$_6$H$_{14}$N$_2$O$_7$ was added to the reaction mixture. The solution was evaporated before the combustion process was initiated. As a result of pyrolysis, accompanied by ignition of the reaction mass, an ultrafine powder with a developed surface and high reactivity is formed, which was calcined at 950 $^{\circ}$ C for 1.5 - 2 hours to remove carbon impurities. The resulting sample was tabletted and calcined in a muffle furnace at 1100 $^{\circ}$C for 8 hours.

\subsection{X-ray diffraction}
X-ray diffraction (XRD) measurements were carried out using the Shimadzu XRD-7000 S automatic diffractometer with exposure 3–5 s in point. X-ray pattern processing was performed with FULLPROF-2020 program.

\subsection{Magnetization measurements}
Magnetization measurements were performed on PPMS-9 device in a temperature range from 13 K to 160 K and in a field range from -10000 to 10000 Oe.

\section{DFT calculations}
To account for self-interaction and strong correlation of electrons in the system, a pseudo-hybrid ACBN0 density functional \cite{acbn0} is employed with norm-conserving pseudopotentials, as implemented in the AFLOW$\pi$ package \cite{aflowpi} integrated with Quantum Espresso code \cite{2009quantum, 2017quantum, 2020quantum}.  This method allows for the self-consistent calculation of the Hubbard $U$ value within the DFT+$U$ approach, possibly with different $U$ values for each crystallographically distinct site. The $U$ values were converged to 0.1 eV accuracy. Perdew-Burke-Ernzerhof (PBE) exchange-correlation functional approximation \cite{pbe} was employed as the basis for DFT+$U$ and ACBN0 calculations. 

We validated our selection of exchange-correlation functional using La$_2$CoO$_4$ as a benchmark -- a compound well-established in the literature as antiferromagnetic (AFM) with a magnetic moment per atom 2.9 $\mu B$ \cite{yamada1989successive}.  In the  ACBN0 approach, AFM ordering emerges as the most favourable magnetic ordering in La$_2$CoO$_4$, yielding a magnetic moment per atom 2.8 $\mu B.$  In contrast, the AFM ordering is not reproduced by various generalized gradient approximation (GGA) functionals such as PBE, RPBE, and PBEsol, all of which predicted the FM ground state. 

A kinetic energy cutoff of 90 Ry and a 6\texttimes6\texttimes6 $k$-point mesh were selected for the calculations. The influence of the energy cutoff was also studied for La$_2$CoO$_4$. Within the energy cutoff range of 90 Ry to 400 Ry, we observed a noteworthy 27\% variation in the energy difference between the ferromagnetic and antiferromagnetic configurations. While a 1 \% discrepancy was achieved at approximately starting from 200 Ry, we could not afford to use this energy cutoff due to the limitations of our computational resources. Throughout the entire energy cutoff range, the AFM configuration consistently exhibited a lower energy. Thus, using the lower cutoff does not alter the conclusions of our work.  

Convergence of $k$-point mesh density was checked in the non-self-consistent field (nscf) calculation for each structure. The difference between the Fermi energies obtained in the nscf field and self-consistent field (scf) calculations is less than 0.01 eV. 

We used scalar-relativistic norm-conserving pseudopotentials from the PSlibrary
1.0.0 for all elements \cite{pseudo}. Pseudopotentials for Sr, Co, Fe O were provided with AFLOW$
\pi$ package, and ld1.x code was used to generate pseodopotential for La. 4p, 3d, 4s orbitals were considered as valence for Co and Fe,  6p, 5d, 6s for La, 5p, 4d, 5s for Sr, and 2p, 2s for O. 

To create the simulation supercells, we constructed 2\texttimes2\texttimes2 supercells from the primitive cell. We consider three possible distributions of La/Sr and Co/Fe within corresponding sublattices. This yields a total of 9 non-equivalent configurations. Atomic models for distributions are shown in Figure \ref{struc}. The choice of the distributions was guided by considerations of charge balance and distribution of valence electrons in the structure, as discussed in the literature \cite{LaSrNiFeO4}. First, we considered a uniform distribution. This distribution satisfies charge balance both between the layers, ensuring the number of valence electrons is the same in each layer, and within the layer, leading to a structure where the closest atom in the A (B) position to La (Co) is Sr (Fe), and vice versa. Next, we considered structures with stripes of La (Co) and Sr (Fe), violating the charge balance within the layer, in order to explore influence of this violation on the electronic and magnetic structure. Finally, we examined structures where layers of La/Sr (Co/Fe) alternate.

Even though this supercell size allows us to explore only a limited number of configurations out of many more possible in a realistic material, it still provides valuable insights on how doping distribution affects magnetic and electronic properties. Direct calculations of larger supercells for such a complex multi-component oxide would require significant computational resources. Also, using machine learning or cluster expansion approaches, which are typically used to explore large configurational spaces of crystalline materials \cite{cluster_expansion}, is highly non-trivial due to varying local spins and complex magnetic interactions, and requires method development that is beyond current study.

We perform the variable cell relaxation for each configuration. The atomic positions are relaxed until the remaining force is below $10^{-3}$~Ha/Bohr, and the total energy difference between two scf cycles was less than $10^{-4}$~Ha. Collinear spin-polarization is taken into account. 

To investigate the impact of La/Sr and Fe/Co distribution on magnetic properties, we employed a model Hamiltonian to assess magnetic interactions in the system. The Hamiltonian incorporates isotropic exchange interactions, denoted as $J_1$ for nearest neighbors and $J_2$ for next-nearest neighbors (as shown in Fig. \ref{magn_struc}):
\be
H=\sum_{\left<i, j\right>} J_1 \textbf{S}_{i} \textbf{S}_j+\sum_{\left<\left<i,j\right>\right>} J_2 \textbf{S}_{i} \textbf{S}_{j},
\label{exchange}
\ee
where $\left<i,j\right>$ corresponds to nearest-neighbour interaction over the square lattice (vertical and horizontal) and $\left<\left<\right>\right>$ represents next-nearest interaction (diagonal). The corresponding exchange interactions are also presented in Figure \ref{magn_struc}. Each pair contributes only once to the overall sum. We assumed that in structures where both nearest and next-nearest interactions are present $J_1\gg J_2$, as it is in LaSrFeO$_4$ \cite{LaSrFeO4_magn, LaSrMnO4}, and considered only $J_1$. 

We calculated the total energy of antiferromagnetic and ferromagnetic configurations of iron for each structure using ACBN0 approximations. Assuming that the difference in total energies is determined only by magnetic interactions and described by the model, we estimated the magnetic interaction for each doping distribution. We did not analyze Co-Co and Fe-Co magnetic exchange interactions. Several attempts to achieve convergence in the scf cycle for different magnetic states of Co within the same distribution by adjusting initial magnetic moments and mixer parameters were made. In some instances, the system transitioned to a lower-energy minimum with a different from starting spin of Co ions, while in other cases the scf cycle did not converge. Thus, we conclude that Co magnetic moment (both orientation and magnitude) is dictated by the doping distribution.
Interestingly, we obtained many metastable electronic states with different magnetic configurations using PBE+$U$ for the tested $U$ values within 1-3 eV  range, but calculations with ACBN0 resulted only in one stable magnetic structure. 

\section{Results and discussions}

\begin{figure}
	\includegraphics[width=1\linewidth]{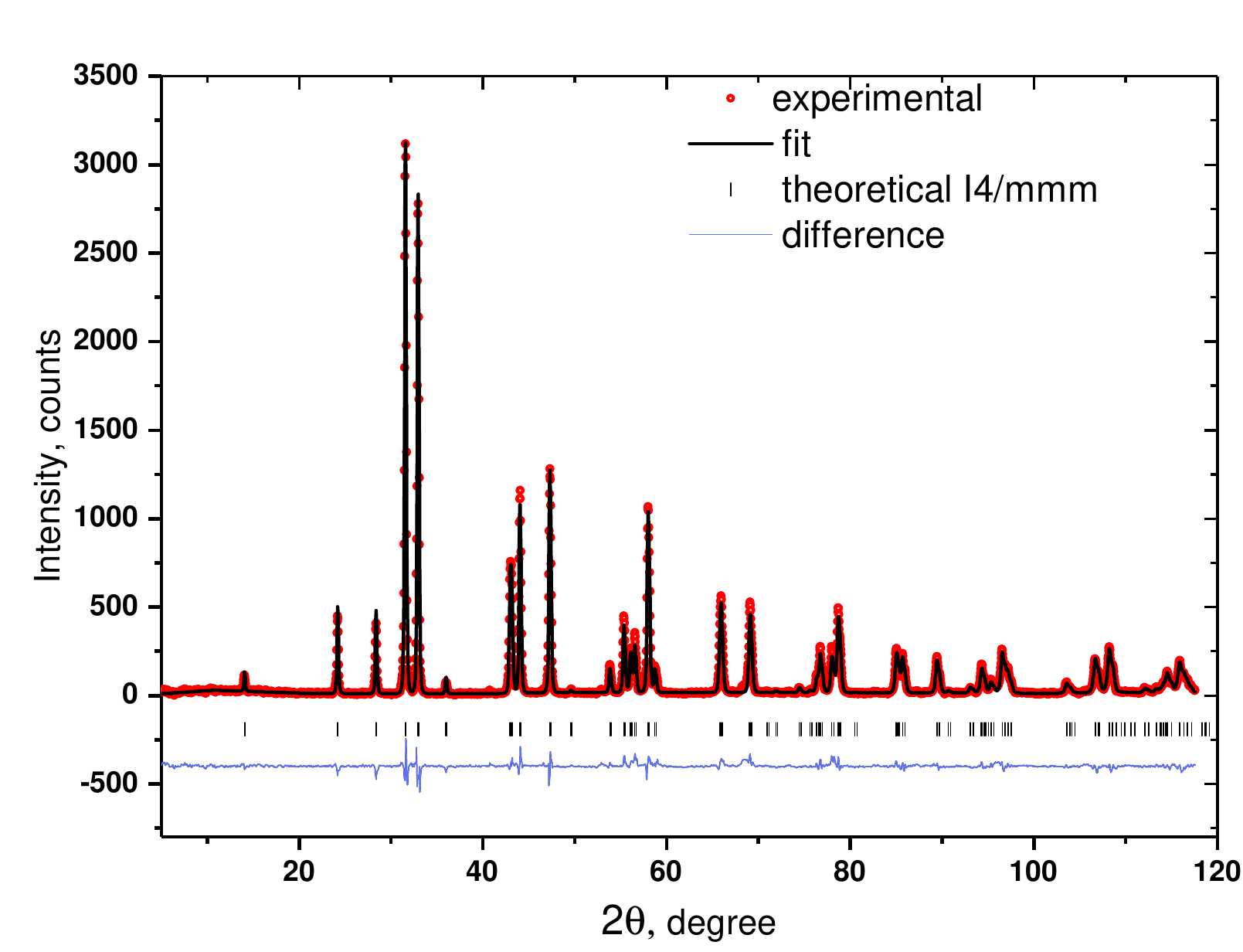} 
    	\caption{The experimental, simulated for I4/mmm  (No 139) crystal with cell parameters derived from experiment, and difference XRD patterns of LaSrCo$_{1/2}$Fe$_{1/2}$O$_4$.}
	\label{struc_ekb_group}
	\end{figure}

XRD pattern for LaSrCo$_{1/2}$Fe$_{1/2}$O$_4$ sample is shown in Fig. \ref{struc_ekb_group}.
The diffraction pattern of the sample is indexed in the space group I4/mmm  (No 139). The structural characteristics derived from the experiment are summarized in Table \ref{struc_table}. All reflections present in the XRD pattern correspond to the K$_2$NiF$_4$-type structure. Impurities were not detected. An assessment of the occupancy of oxygen atom positions, indicating the oxygen stoichiometry of the sample, was also performed and is given in Table \ref{struc_table}. The corresponding R-factors are: weighted profile $R_{\text{wp}} = 9.53\%$, expected $R_{\text{exp}} = 8.54\%$, Bragg $R_{\text{B}} = 4.56\%$, and $\chi^2 = R_{\text{wp}}^2 / R_{\text{exp}}^2 = 1.24$.

We also attempted to identify the diffraction pattern within the I4mm structure (No. 107) corresponding to the arrangement where layers of La and Sr are interchanged, as this configuration is the most thermodynamically stable according to the DFT results (see below). However, the corresponding $\chi^2$ value was 2.16, and refining the occupancies of La and Sr resulted in distributions of 0.47 vs. 0.53 in one layer and opposite in the other. These findings likely suggest the absence of La/Sr layer ordering, although this conclusion should be approached with caution.

Superlattice peaks indicating orderings in A or B positions appear only if these orderings cause lattice distortions or reduce symmetry due to different atomic positions, rather than just different occupancies. For example, in \cite{La4LiMnO8}, the ordering in the B position for ions with significantly different masses, such as Li and Mn, was not established using XRD or even neutron diffraction, but was detected using local probe techniques such as nuclear magnetic resonance.

The comparison of calculated and experimental lattice parameters and interatomic distances with the literature data are presented in Table \ref{struc_comparison}. The values of the unit cell parameters are in satisfactory agreement with the literature data \cite{synthesis3}. However, the LaSrCo$_{1/2}$Fe$_{1/2}$O$_4$ sample under study is characterized by a more pronounced anisotropy of the coordination polyhedra. (Co/Fe)O$_6$ octahedron is elongated along the c axis, and the (La/Sr)O$_9$ antiprism is compressed. The interatomic distances La/Sr - O1, La/Sr - O2, and Co/Fe - O1 are comparable to the values of interatomic distances in LaSrCo$_{1/2}$Fe$_{1/2}$O$_4$ obtained by the solid-phase method at 1350 $^{\circ}$ C \cite{synthesis1}.

\begin{table}[]
\begin{tabular}{lllllll}
Atom  & Site & x & y   & z          & $B_{\text{iso}}$   $\AA^2$ & Occupancy \\
La/Sr & 4e   & 0 & 0   & 0.3598(5) & 0.588(9)    & 1         \\
Co/Fe & 2a   & 0 & 0   & 0          & 0.64(3)     & 1       \\
O1    & 4e   & 0 & 0.5 & 0          & 1.79(8)     & 1.03(4)   \\
O2    & 4c   & 0 & 0   & 0.1675(6) & 3.25(5)     & 0.99(4)  
\end{tabular}
\caption{Structural characteristics of the RP oxide LaSrCo$_{1/2}$Fe$_{1/2}$O$_4$. See Fig. \ref{magn_struc} for the site labelling description.}
\label{struc_table}
\end{table}

The ACBN0 lattice parameters vary depending on the doping distribution and magnetic state. The computed lattice parameter $c$ varies between 13.2 and 14 $\AA$. As expected, structures with the same amount of La and Sr in each layer have a larger $c$ than structures in which the layers of La and Sr are interchanged. This is due to the difference in ionic radii of La and Sr. The second factor that controls $c$ parameter is the Co spin state. We find that structures in which Co is in a low-spin state have a smaller $c$ parameter. The values of $a$ and $b$ parameters vary between 3.85 $\AA$ and 3.925 $\AA$. This variation is much smaller than the variation of the $c$ parameter. The highest ratio of $b/a$ is 1.008.
\begin{table*}[!htb]
	\begin{tabular}{lllllllll}
		& exp & min & max &LSCFO1   & LSCFO2            & LSCO & LSFO &           LCFO         \\
		a               & 3.8382    & 3.85      & 3.925     & 3.83 & 3.823                                         & 3.798                                           & 3.87                                             &                                                    \\
		c        &  12.558   & 13.2     & 14    & 12.52  & 12.964 &12.525                                          & 12.71                                            &                                                    \\
		Co-OI  & 1.919$^*$    & 1.872    & 1.975    &                                              & 1.92$^*$ & 1.902     &                                                  & 1.981  \\
		Co-OII     & 2.107$^*$   & 2.066    & 2.348    &                                              &2.086$^*$ & 1.906                                           &                                                  & 2.312                                              \\
		Fe-OI      &  1.919$^*$   & 1.868    & 1.975    &                                              &  1.92$^*$&                                                & 1.937      & 2.002                                              \\
		Fe-OII & 2.106$^*$  & 2.149    & 2.419    &                                              &    2.086$^*$    &                                         & 2.127      & 2.294                                             
	\end{tabular}

$^*$averaged value among Fe-O and Co-O distance 
\caption{Comparison of lattice parameters, transition metal--oxygen bond lengths in the obtained sample, and minimum (min) and maximum (max) ACBN0 computed values among different distributions of La/Sr and Fe/Co and magnetic configurations with the literature data:  LaSrFe$_{1/2}$Co$_{1/2}$O$_4$ \cite{small_polaron} (LSCFO1),  LaSrFe$_{1/2}$Co$_{1/2}$O$_4$ \cite{synthesis1} (LSCFO2),  LaSrCoO$_4$ \cite{doping_1-1.4} (LSCO), LaSrFeO$_4$\cite{LaSrFeO4_magn_sus} (LSFO),La$_2$Co$_{1/2}$Fe$_{1/2}$O$_4$ \cite{La2Fe0.5Co0.5O4first} (LCFO). The data are presented for conventional lattice unit cell. All the distances are measured in angstroms.}
\label{struc_comparison}
\end{table*}

To check whether magnetic interaction is present between the layers, we calculated the energy difference between the FM and AFM-A type magnetic structures. In AFM-A type magnetic structures magnetic moments in different layers are aligned in opposite directions, whereas within the same layer they are aligned in the same direction as presented in Figure \ref{magn_struc}. Both FM and AFM-A structures were fully relaxed. The resulting energy difference is of the order of $10^{-5}$ eV per primitive unit cell. This value is much smaller than the calculated energy differences for structures with different distributions of elements and magnetic configurations within the same layer. Therefore, the magnetic structure can be considered 2D, as is commonly accepted for RP phases with $n$ = 1, except for some exotic cases.

\begin{figure*}
	\includegraphics[width=0.47\linewidth]{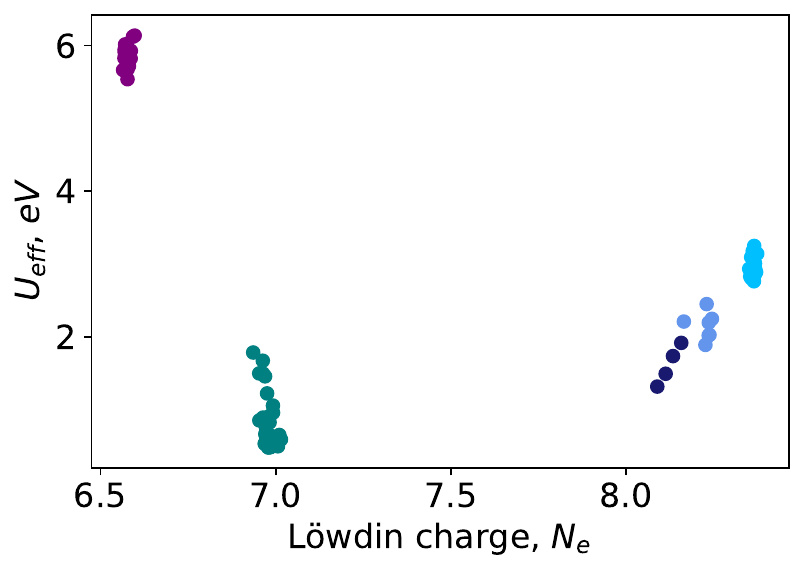}
	\includegraphics[width=0.47\linewidth]{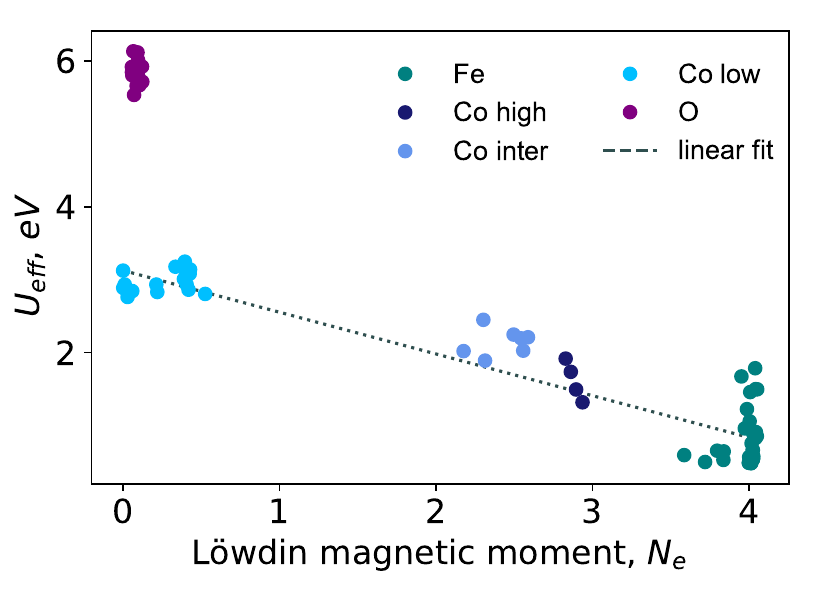}
	
	\caption{The correlation between converged $U$ value and charge (left) or magnetic moment (right). Purple color represents oxygen, teal--iron, midnightblue -- cobalt in high-spin state, cornflowerblue -- cobalt in intermediate spin state, and deepskyblue -- cobalt in low-spin state. In metallic structures L{"o}wdin magnetic moment greater than 2.7 assigned as high spin state and lower than 2.7 but greater than 2 as intermediate spin state.    }
	\label{U}
\end{figure*}

To identify Co and Fe spin states, we calculated L{\"o}wdin charges and magnetic moments (Figure \ref{U}). Upon initial examination, Co appears to exist in two distinct spin states: (i) a high-spin state with L{\"o}wdin magnetic moments ranging from 2.15 to 3 $\mu_B$, and (ii) a low-spin state with L{\"o}wdin  magnetic moments ranging from 0 to 0.6 $\mu_B$. Magnetic moment of iron varies within a range of 3.5 to 4 $\mu_B$. These values are quite peculiar. First, there is a noticeable variation of magnetic moments even within each spin state. Second, the range of 2.15 to 3 $\mu_B$ for Co$^{3+}$ ion could potentially correspond to both intermediate- and high-spin states, since Co$^{3+}$ formally possesses 6 electrons in the 3$d$ orbitals and can exhibit a low-spin state with zero unpaired electrons, an intermediate-spin state with two unpaired electrons, and a high-spin state with four unpaired electrons. Thus, L{\"o}wdin partitioning analysis indicates a discrepancy between the simplified pure-ionic picture and real material.

\begin{figure}
	\includegraphics[width=1\linewidth]{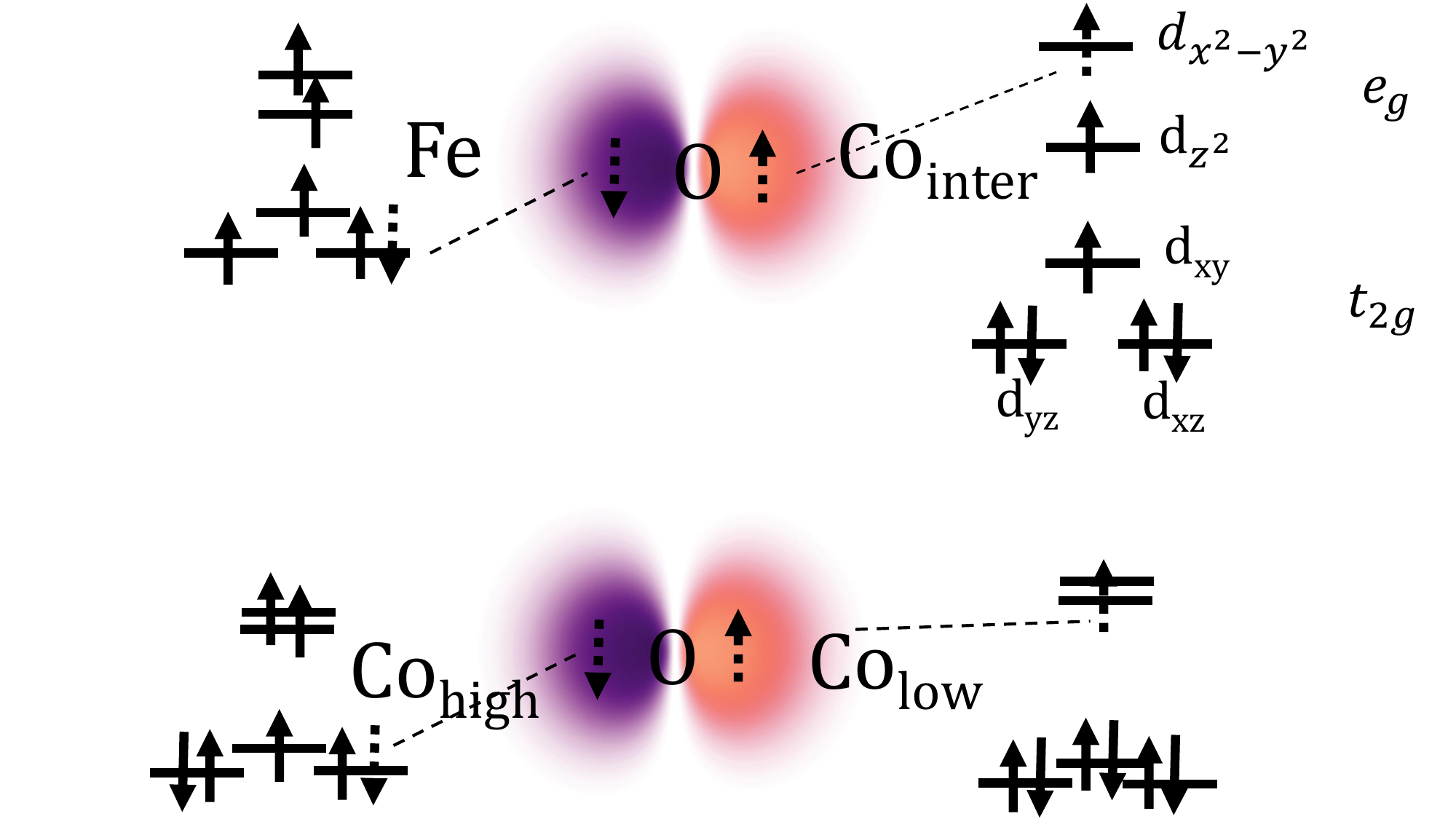}	
	\caption{Illustration of partial charge and magnetic moment delocalization between oxygen and transition-metal ions according to the L{\"o}wdin partitioning. Different Co$^{3+}$ spin states in octachedral environment are shown. }
 \label{Co_spin_states}
\end{figure}

The variation of the magnetic moments is explained by delocalization of electron density between the metal and oxygen ions. Due to this delocalization, L{\"o}wdin partitioning assigns some of the oxygen electrons in one of the spin channels to the neighboring transition-metal ion, as shown graphically in Figure \ref{Co_spin_states}. The spin channel of the "additional" electron density may be different for Co in intermediate- and low-spin states depending on the next-neighbor transition-metal spin.

In structures with an integer total number of unpaired electrons per unit cell (typically with a finite electronic band gap), magnetic moments can be analyzed in more detail. When magnetization is only due to Fe, while Co is in low-spin state, the total magnetic moment corresponds to Fe$^{3+}$ in a high-spin state. In most other structures with an integer total number of unpaired electrons per unit cell, the L{\"o}wdin magnetic moment of Fe and Co are close to the middle value between high-spin and intermediate-spin states ($\sim$4 for Fe and $\sim$3 for Co), and it is therefore not possible to assign unambiguously the spin states. Assuming that Fe is in a high-spin state, since Fe$^{3+}$ in a low or intermediate state is rare, Co should be in an intermediate-spin state in these structures, except in one structure where Co is in a high-spin state according to the total magnetic moment of the unit cell.  In metallic structures, it is not possible to identify a particular spin state of TM ions since electrons are delocalized and even the total number of unpaired electrons obtained in calculations in the unit cell is not integer. 

\begin{figure}
	\includegraphics[width=1\linewidth]{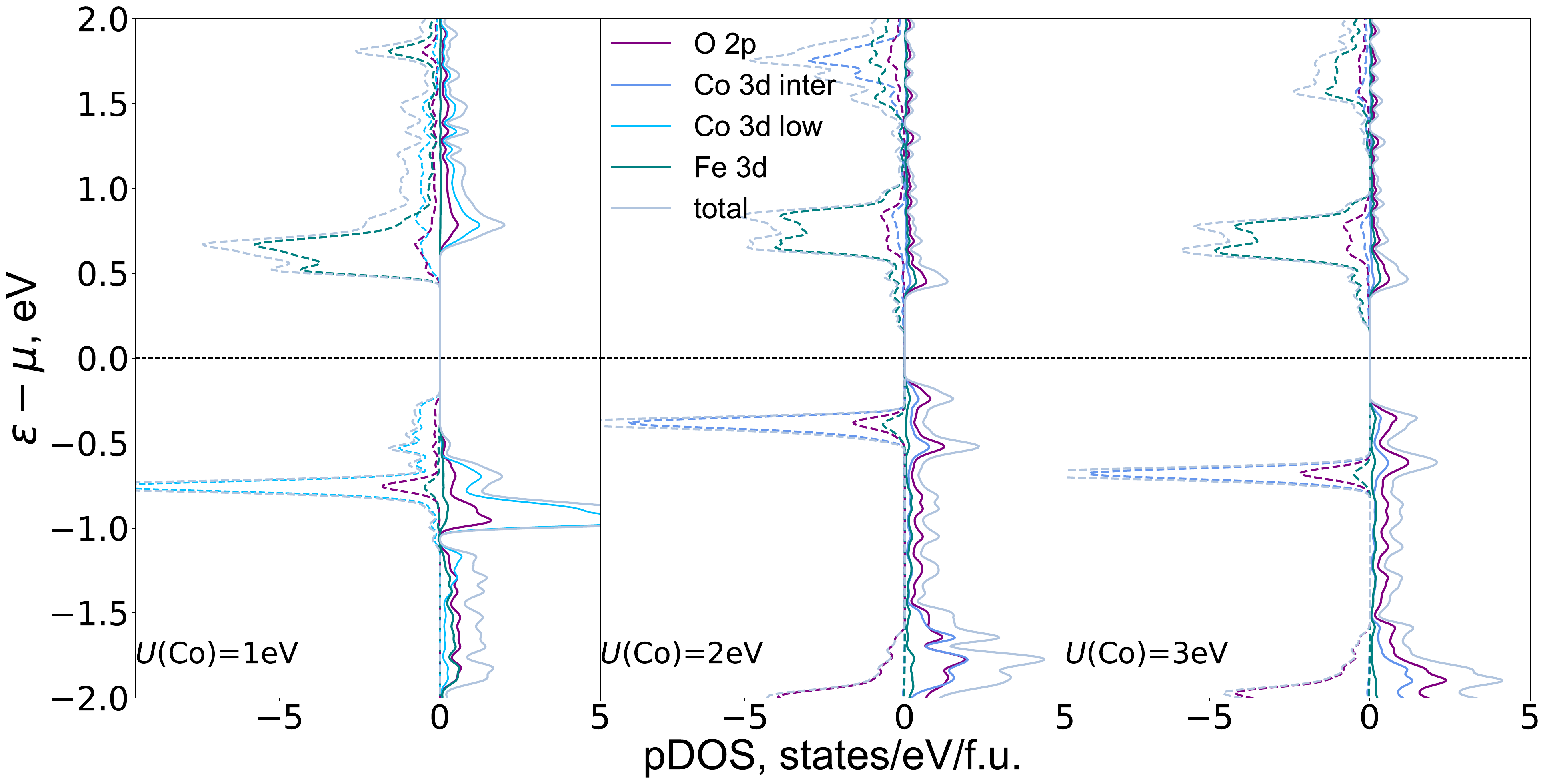}
	\caption{Calculated DOS of the most energetically favorable structure (structure (a) in Fig \ref{struc}) with different $U$ for Co. Purple color represents oxygen, teal--iron, blue--cobalt in high spin state and dodgerblue--cobalt in low spin state, grey -- total DOS. Dashed line corresponds to the spin minority states.}
	\label{diff_U}
\end{figure}

The converged $U$ values are shown in Figure \ref{U}. Notably, we observe a variation of $U$ by up to $1.5$ eV for the same species within the same spin state. To understand the impact of these $\pm1$ eV variations on the electronic and magnetic properties within our system, we conducted density-of-states (DOS) calculations for a structure where layers of La/Sr are interchanged, and Co and Fe are uniformly distributed (Fig. \ref{struc} a), utilizing different $U$ values for Co. The results are shown in Figure \ref{diff_U}. First, as anticipated, the band gap changes with varying $U$. Increasing the ACBN0 self-consistent $U$ value of 2 eV to 3 eV results in a band gap increase from 0.14 eV to 0.3 eV. Second, adjusting $U$ to 1 eV not only alters the band gap but also changes Co spin state from intermediate to low. We have identified a good correlation between the $U$ value for the transition-metal $d$ states and its magnetic moment with the coefficient of determination $R^2=0.87$.

Our calculations reveal that for any distribution on the B-site the most energetically favorable arrangement of La and Sr is a structure, where layers of La alternate with layers of Sr (Fig. \ref{struc} a-c). This ordering is observed in certain perovskites and Ruddlesden-Popper phases \cite{ordered_disordered_perovskite, lay_ordering}. For this A-site ordering, the structure that maximizes the number of Co-O-Fe bridges and exhibits ferromagnetic ordering is the most favorable among all the B-site arrangements. Its energy is lower by at least 0.16 eV per primitive unit cell than the energies of other distributions. This is the only structure among the ones considered in our study where iron exhibits ferromagnetic ordering. We note that metastable configurations can be present in synthesized material due to kinetic limitations, as indicated by the comparison of experimental and theoretical magnetic properties discussed below.

\begin{figure}

	\includegraphics[width=1\linewidth]{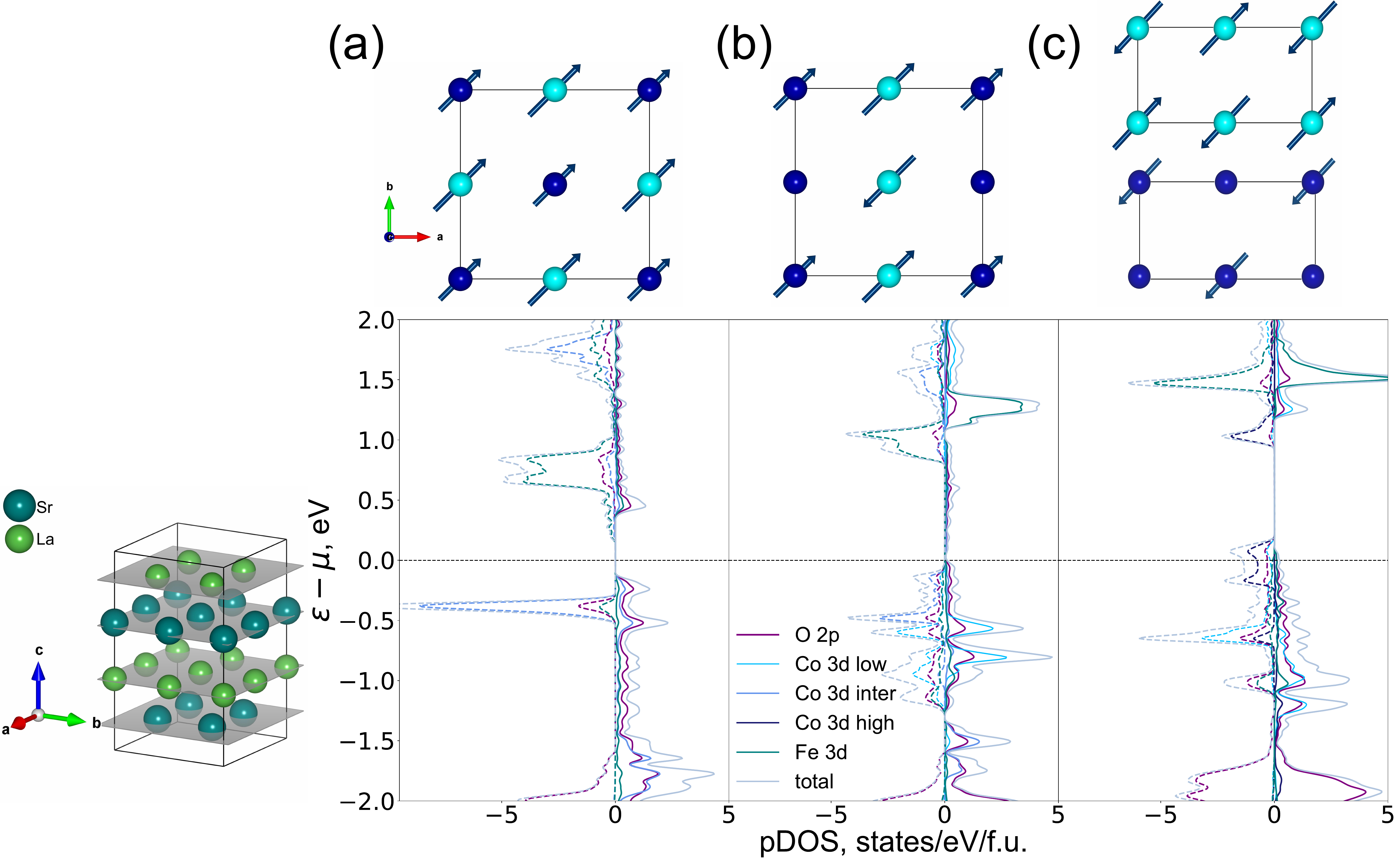}
 
	\includegraphics[width=1\linewidth]{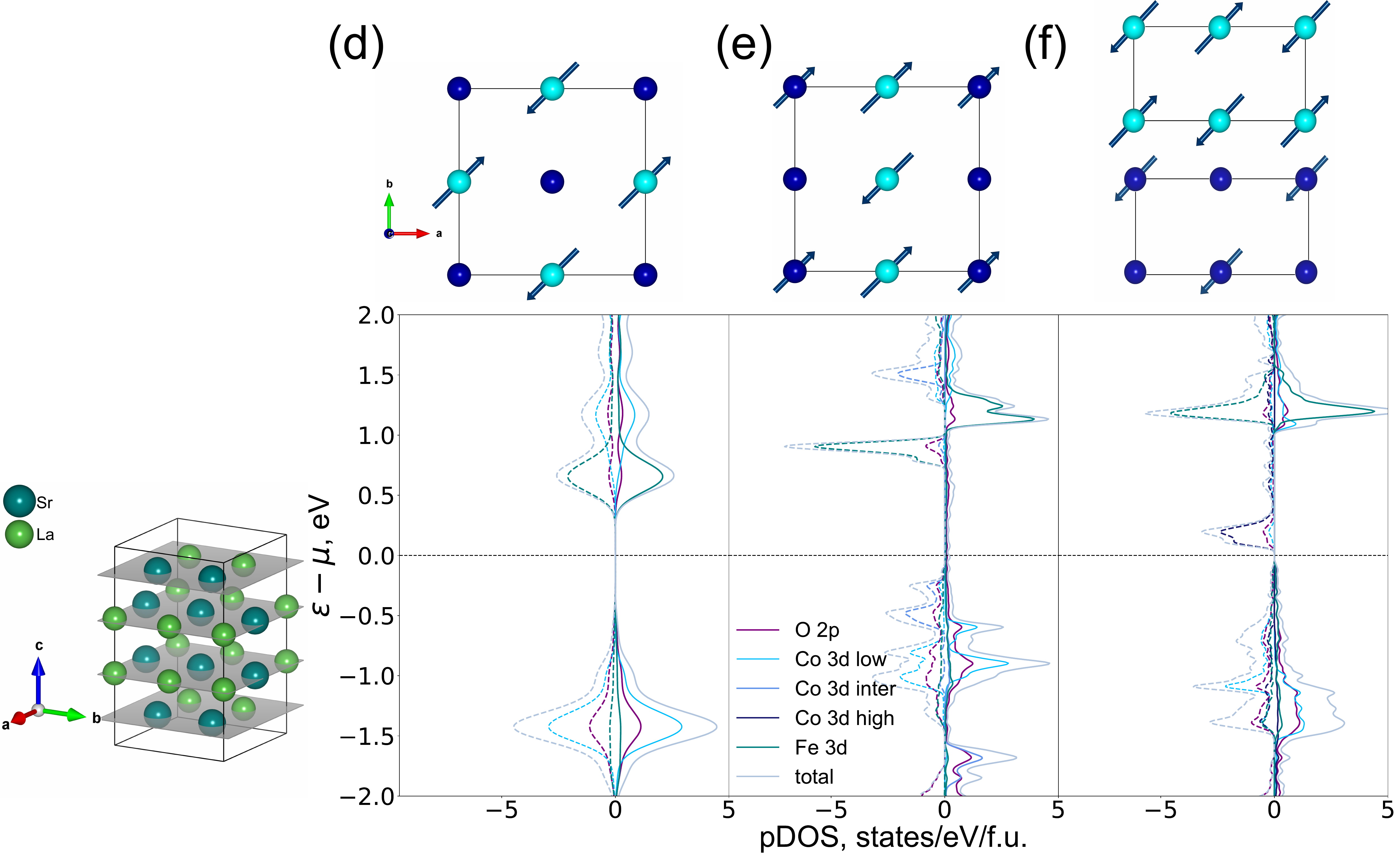}
 
	\includegraphics[width=1\linewidth]{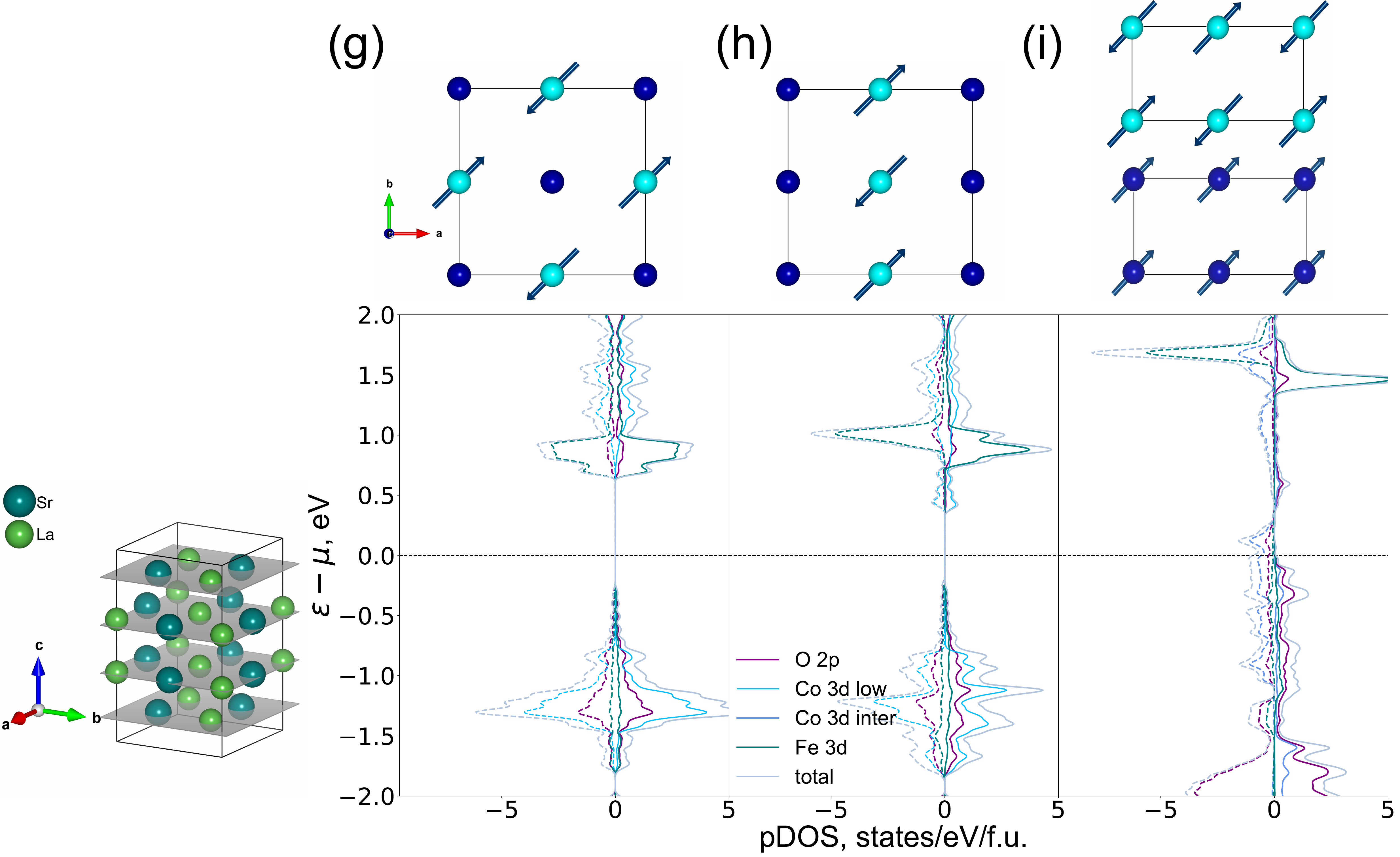}
	
	\caption{Projected DOS and most favorable magnetic orderings of various Fe/Co and La/Sr distributions in LaSrCo$_{1/2}$Fe$_{1/2}$O$_4$.
  A top view of the Co/Fe distribution with the most favorable magnetic arrangement is presented in each figure. In structures where layers of Co and Fe are interchanged, both layers are shown. The corresponding distribution of La/Sr is shown in each figure on the left side. Only La/Sr is displayed, with grey slices serving as guides for the eyes to identify the same layers of La/Sr.  The structure models are presented in Fig. \ref{struc} in the same order.  Purple color represents oxygen, teal--iron, blue--cobalt in high spin state and dodgerblue--cobalt in low spin state grey -- total DOS. Dashed line corresponds to the spin minority states.}
	\label{DOS}
\end{figure}

The computed projected DOS and energetically preferred magnetic orderings are presented in Figure \ref{DOS}. In an octahedral environment, the 3$d$ orbitals of iron and cobalt undergo a crystal field splitting into two sets of degenerate states: the upper $e_g$ states and the lower $t_{2g}$ states as shown in Figure \ref{Co_spin_states}. If the energy splitting $\delta$ is sufficiently large, electrons occupy the lower energy $t_{2g}$ orbitals before populating the higher energy $e_g$ orbitals. The degeneracy is further lifted due to the elongation and tilting of the O-octahedra. 

The projected DOS reveals a strong hybridization between O 2$p$ and metal 3$d$ orbitals in each structure. For instance, changes in the transition-metal's magnetic configuration lead to alterations in the DOS projected on O 2$p$ orbitals. This strong hybridization is consistent with the delocalization of magnetic moments between O and metal ions discussed above (see Figure \ref{Co_spin_states}).

Several trends and similarities in electronic structure of different distributions in A- and B-sites can be readily identified based on our ACBN0 calculations. First, the states around the Fermi level are formed by oxygen's 2$p$ states and the 3$d$ states of the transition metals. Second, the unoccupied Fe 3$d$ states are localized in each configuration. Third, with reducing the number of Fe-O-Co bridges, the 3$d$ Co orbitals become delocalized. Fourth, occupied Co low-spin states are localized. The primary difference arises from the Co high-spin states. Despite the similarities in the DOS, key features such as the band gap differ from one distribution to another. Since the value and nature of DOS around the Fermi level is important for various applications (e.g., for electrocatalytic oxidation the unoccupied states accommodate the holes that can participate in the oxidation), we analyze the influence of atomic configuration on these properties. Structures in which Co and Fe are uniformly distributed (Fig.\ref{DOS}, a,d,g) exhibit semiconductor behavior, with states near conduction band minimum (CBm) being primarily composed of iron 3d orbitals, and the states near valence band maximum (VBM) being composed of cobalt 3d orbitals. In structures where Co and Fe are arranged in stripes (Fig. \ref{DOS} b, e, h), the states around the Fermi level are determined by the Co spin states. When Co in a high-spin state interchanges with Co in a low-spin state (Fig. \ref{DOS} a,b), the material is a semi-metal: there is no band gap in the spin-minority channel. In this channel, states at and around the Fermi level are formed by Co 3$d$ and O 2$p$ states, indicating a significant covalent character of the Co-O bonding and the mixed character of the electronic holes. When Co is in  low-spin states, the electronic structure closely resembles that of uniformly distributed Co and Fe structures: a band gap exists, with states near CBm primarily formed by Fe 3$d$ states, and the states near VBM composed by Co 3$d$ orbitals. In structures where layers of Co and Fe are interchanged (Fig. \ref{DOS} c, f, i), states near the Fermi level are formed by Co 3$d$ states. Structures with uniformly distributed La/Sr (Fig. \ref{DOS} c) and  layers of La/Sr interchanged (Fig. \ref{DOS} i)  exhibit metallic behavior, whereas the structure where La/Sr ordered in stripes (Fig. \ref{DOS} f) is a semiconductor.

\begin{table}[]
	\begin{tabular}{llllll}
		& min & max & LaSrFeO$_4$ & LaSrFeO$_4$  &LaSrMnO$_4$   \\
		J$_1,$ meV & 10  & 37& 19/29&   7.4                                         &                       3.4                         \\
		J$_2,$ meV & 8  &11   & & 0.4                                          & 0.4                                               
	\end{tabular}
\caption{Comparison of exchange coupling coefficients (minimum and maximum among all calculated values) in LaSrCo$_{1/2}$Fe$_{1/2}$O$_4$ with calculated values for two arrangements of La/Sr in LaSrFeO$_4$ and  literature values for LaSrFeO$_4$ \cite{LaSrFeO4_magn} and LaSrMnO$_4$ \cite{LaSrMnO4}. The coefficients are obtained by fitting Eq. \ref{exchange} to ACBNO data, as described in the text.}
\label{exchange_table}
\end{table}

Magnetic moments of Fe cations are ordered antiferromagnetically in eight out of the nine considered structures (see Fig. \ref{DOS}). Among these, six structures have integer total number of unpaired electrons per unit cell. Therefore, spin state of Co in these configurations can be identified clearly: one structure features Co in both high-spin state and low-spin state (\ref{struc} f), two structures include a mixture of intermediate-spin state and low-spin state (\ref{struc} e and b), and in three structures Co is in a low-spin state (\ref{struc} d, g, and h). In the two metallic structures, free-ion spin states cannot be assigned to Co ions because the total number of unpaired electrons corresponds to the spin state between high and intermediate values. In the ferromagnetically ordered structure (\ref{struc} a) Co has intermediate-spin state. 

These results align with literature data for LaSrCoO$_4$ and LaSrFeO$_4$. In the study conducted by Haw {\em et al}., a composition of 40\% high-spin and 60\% low-spin configurations was observed in LaSrCoO$_4$ \cite{high_low_exp}. {\em Ab initio} calculations supported this observation, showing a 50-50\% mixture of these configurations \cite{high_spin_low_spin}. This finding was further corroborated by Hartree-Fock restricted calculations \cite{high_low_HF_calc}. LaSrFeO$_4$ is recognized for exhibiting antiferromagnetic ordering with estimated Neel temperature $T_N$ = 366 K according to neutron diffraction and M{\"o}ssbauer spectrum measurements. However, susceptibility measurements suggest a slightly lower value $T_N$ = 325 K \cite{LaSrFeO4_magn_sus}. The antiferromagnetic coupling coefficients $J_1=7.4$ meV between nearest neighbors and $J_2=0.4$ meV between next-nearest neighbors are reported\cite{LaSrFeO4_magn}. 

Calculated exchange interactions in LaSrCo$_{1/2}$Fe$_{1/2}$O$_4$ using the formula \ref{exchange} are presented in table \ref{exchange_table}.  We stress that exchange interaction parameters are very sensitive to the calculation method as well as to the inter-atomic distances \cite{exchange_comparison, exchange_comparison2}. To provide a reference value for our results, we calculated exchange interaction parameters for LaSrFeO$_4$ for uniform distribution of La/Sr and for the structure where layers of La interchanged with layers of Sr. The calculated values are also presented in table \ref{exchange_table}. The magnitude and variation range of these values are of the same order as for LaSrCo$_{1/2}$Fe$_{1/2}$O$_4$. The difference between the maximum and minimum calculated values due to variations in doping distribution is at least comparable to the difference arising from the transition-metal substitution. 

\begin{figure}
	\includegraphics[width=1\linewidth]{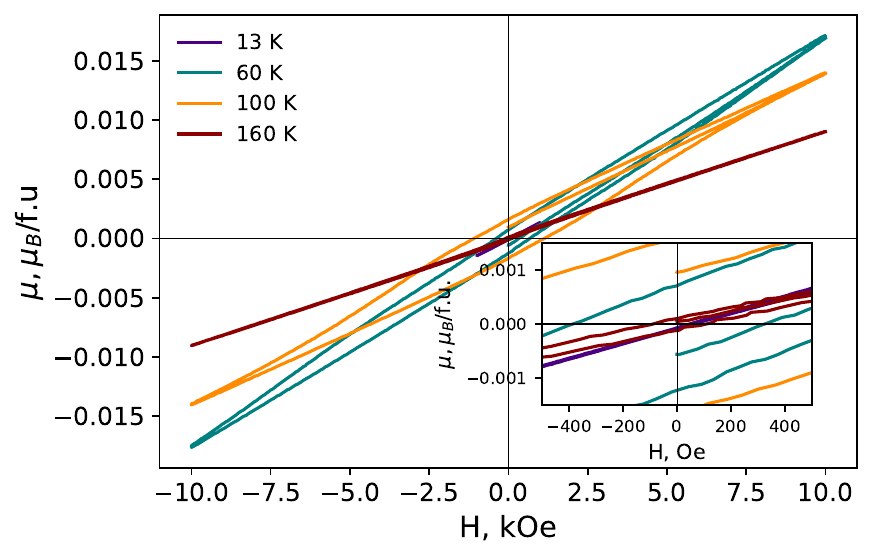}	
	\caption{Magnetization as a function of the applied field at various temperatures (hysteresis). At 28 K and below, no hysteresis is observed. The character of magnetization suggests AFM ordering.}
 \label{loops}
\end{figure}

To determine whether LaSrCo$_{1/2}$Fe$_{1/2}$O$_4$ is ferromagnetic, ferrimagnetic or antiferromagnetic, we conducted a series of magnetization measurements. The results are presented in Figure \ref{loops}. The first faint hysteresis loop appears at 160 K. As expected, the saturation magnetization decreased with increasing temperature. However, the coercive filed $H_c$ and remnant magnetization $M_r$ depend non-monotonically on temperature reaching their highest values at $T$ = 100 K, with $H_c$ = 1040 Oe and $\mu_r$ = 1.6 $\times 10^{-3} \mu_B$/f.u., respectively. We attribute the disappearance of hysteresis at low temperatures to the transition of cobalt from a high-spin state to a low-spin state as the temperature decreases. According to our calculations, the lowest-energy distribution of dopants and spins is ferromagnetic (Fig. \ref{struc} a). However, real samples can contain a metastable distribution, and most of the metastable distributions that we calculated are antiferromagnetic. Thus, comparison between theory and experiment indicates that synthesized samples do not have the thermodynamically stable distribution of dopants.

A minor exchange bias was observed at $T$ = 60 K, with coercive field $H_c$ = -425 Oe for negative field and $H_c$ =693 Oe for positive field, and remnant magnetization $M_r$ = $-12 \times 10^{-4} \mu_B$/f.u. for negative and $M_r$ = $7 \times 10^{-4} \mu_B$/f.u for positive field, respectively. The negligibly small magnitude of the experimental magnetic moment suggests the absence of ferrimagnetic coupling between iron and cobalt. However from the reentrant behavior of the coercive force and remnant magnetization one can conclude that magnetization loops have internal nature of the sample and exclude an ordinary ferromagnetic impurity, for which both coercive force and remnant magnetization would increase with decreasing temperature.  

\section{Conclusions}
In our study of the Ruddlesden-Popper oxide LaSrCo$_{1/2}$Fe$_{1/2}$O$_4$ we investigated the impact of doping distribution on its key properties. The distribution of La/Sr played a primary role in determining lattice parameters due to the appreciable difference in their ionic radii. Additionally, the Co-spin states emerged as a significant factor, with low-spin state cobalt aligning with smaller distances between cobalt and surrounding oxygen atoms.

 Both A-site and B-site distributions strongly influence magnetic and electronic properties. We attribute the variation in DOS for different La/Sr and Co/Fe distributions mainly to the delocalized states of Co in the high-spin state, whereas localized states of Co in low-spin state and Fe change insignificantly from configuration to configuration. Comparison of the calculations and experiment shows that magnetic ordering is not solely determined by the most thermodynamically stable distribution but by a mixture of distributions. A subtle hysteresis was observed in the temperature range of 60-160 K. At very low temperatures hysteresis loops disappeared, indicating the transition of cobalt to a low-spin state. These findings reveal a subtle interplay between doping distribution and the electronic properties of LaSrCo$_{1/2}$Fe$_{1/2}$O$_4$, which is important for design of functional materials based on this and other mixed-metal RP oxides.
 
\section{Acknowledgment}
This work was supported by the Russian Science Foundation under grant no. 21-13-00419
\bibliography{LaSrCo05Fe05O4}

\end{document}